\documentclass[journal]{IEEEtran}
\usepackage{graphicx}
\usepackage{amsmath}
\usepackage{subfigure}
\usepackage{rotating}
\usepackage{multirow}
\usepackage{array}
\usepackage{rotating}

\begin{document}
\title{On Performance Evaluation of\\ Variants of DEEC in WSNs}

\author{T. N. Qureshi, N. Javaid, M. Malik, U. Qasim$^{\ddag}$, Z. A. Khan$^{\S}$\\

        COMSATS Institute of IT, Islamabad, Pakistan. \\
        $^{\ddag}$University of Alberta, Alberta, Canada\\
        $^{\S}$Faculty of Engineering, Dalhousie University, Halifax, Canada.
        }

\maketitle

\begin{abstract}
Wireless Sensor Networks (WSNs) contain numerous sensor nodes having limited power resource, which report sensed data to the Base Station (BS) that requires high energy usage. Many routing protocols have been proposed in this regard achieving energy efficiency in heterogeneous scenarios. However, every protocol is not suitable for heterogeneous WSNs. Efficiency of protocol degrades while changing the heterogeneity parameters. In this paper, we first test Distributed Energy-Efficient Clustering (DEEC), Developed DEEC (DDEEC), Enhanced DEEC (EDEEC) and Threshold DEEC (TDEEC) under several different  scenarios containing high level heterogeneity to low level heterogeneity. We observe thoroughly regarding the performance based on stability period, network life time and throughput. EDEEC and TDEEC perform better in all heterogeneous scenarios containing variable heterogeneity in terms of life time, however TDEEC is best of all for the stability period of the network. However, the performance of DEEC and DDEEC is highly effected by changing the heterogeneity parameters of the network.
\end{abstract}

\begin{IEEEkeywords}
Cluster, Head, Residual, Energy, Heterogenous, Efficient, Wireless, Sensor, Networks
\end{IEEEkeywords}

\section{Introduction}
Technological developments in the field of Micro Electro Mechanical Sensors (MEMS) have enabled the development to tiny, low power, low cost sensors having limited processing, wireless communication and energy resource capabilities. With the passage of time researchers have found new applications of WSN. In many critical applications WSNs are very useful such as military surveillance, environmental, traffic, temperature, pressure, vibration monitoring and disaster areas. To achieve fault tolerance, WSN consists of hundreds or even thousands of sensors randomly deployed inside the area of interest [1]. All the nodes have to send their data towards BS often called as sink. Usually nodes in WSN are power constrained due to limited battery, it is also not possible to recharge or replace battery of already deployed nodes and nodes might be placed where they can not be accessed. Nodes may be present far away from BS so direct communication is not feasible due to limited battery as direct communication requires high energy. Clustering is the key technique for decreasing battery consumption in which members of the cluster select a Cluster Head (CH). Many clustering protocols are designed in this regard [2,3]. All the nodes belonging to cluster send their data to CH, where, CH aggregates data and sends the aggregated data to BS [4-6]. Under aggregation, fewer messages are sent to BS and only few nodes have to transmit over large distance,so high energy is saved and over all lifetime of the network is prolonged. Energy consumption for aggregation of data is much less as compared to energy used in data transmission. Clustering can be done in two types of networks i.e homogenous and heterogeneous networks. Nodes having same energy level are called homogenous network and nodes having different energy levels called heterogeneous network. Low-Energy Adaptive Clustering Hierarchy (LEACH) [5], Power Efficient GAthering in Sensor Information Systems (PEGASIS) [7], Hybrid Energy-Efficient Distributed clustering (HEED) [8] are algorithms designed for homogenous WSN under consideration so these protocols do not work efficiently under heterogeneous scenarios because these algorithms are unable to treat nodes differently in terms of their energy. Whereas, Stable Election Protocol (SEP) [9], Distributed Energy-Efficient Clustering (DEEC) [10], Developed DEEC (DDEEC) [11], Enhanced DEEC (EDEEC) [12] and Threshold DEEC (TDEEC) [13] are algorithms designed for heterogeneous WSN. SEP is designed for two level heterogeneous networks, so it can not work efficiently in three or multilevel heterogeneous network. SEP considers only normal and advanced nodes where normal nodes have low energy level and advanced nodes have high energy. DEEC, DDEEC, EDEEC and TDEEC are designed for multilevel heterogeneous networks and can also perform efficiently in two level heterogeneous scenarios.

In this paper, we study performance of heterogeneous WSN protocols under three and multi level heterogeneous networks. We compare performance of DEEC, DDEEC, EDEEC and TDEEC for different scenarios of three and multilevel heterogeneous WSNs. Three level heterogeneous networks contain normal, advanced and super nodes whereas super nodes have highest energy level as compared to normal and advanced nodes.

We discriminate each protocol on the basis of prolonging stability period, network life time of nodes alive during rounds for numerous three level heterogeneous networks. Each containing different ratio of normal, advanced and super nodes along with the multilevel heterogeneous WSNs.

It is found that different protocols have different efficiency for three level and multilevel heterogeneous WSNs in terms of stability period, nodes alive and network life time. DEEC and DDEEC perform well under three level heterogeneous WSNs containing high energy level difference between normal, advanced and super nodes in terms of stability period. However, it lacks in performance as compared to EDEEC and TDEEC in terms of network lifetime. Whereas, EDEEC and TDEEC perform well under multi and three level heterogeneous WSNs containing low energy level difference between normal, advanced and super nodes in terms of both stability period and network lifetime.
\section{Related Work}
Heinzeman, \textit{et al.} [5] introduced a clustering algorithm for homogeneous WSNs called as LEACH in which nodes randomly select themselves to be CHs and pass on this selection criteria over the entire network to distribute energy load. G. Smaragdakis, \textit{et al.} [9] proposed a protocol called as SEP in which every sensor node in a heterogeneous two level hierarchical network independently elects itself as a CH based on its initial energy relative to other nodes. L .Qing, Q. Zhu and M. Wang [10] worked on heterogeneous WSN and proposed a protocol named as DEEC in which CH selection is based on the basis of probability of the ratio of residual energy and average energy of the network. Brahim Elbhiri, \textit{et al.} [11] worked on heterogeneous WSN and proposed a protocol named as DDEEC  is based on residual energy for CH selection to balance it over the entire network. So, the advanced nodes are more likely to be selected as CH for the first transmission rounds, and when their energy decreases, these nodes will have the same CH election probability like the normal nodes. P. Saini \textit{et al.} [12] proposed  a protocol EDEEC which is extended to three level heterogeneity by adding an extra amount of energy level known as super nodes. Parul Saini and Ajay K Sharma [13] proposed  a protocol TDEEC scheme selects the CH from the high energy nodes improving energy efficiency and lifetime of the network.

\section{Motivation}
Many algorithms are recently proposed to increase stability and lifetime of heterogeneous WSNs. However, heterogeneous networks are of different types having different parameters. Every algorithm does not work efficiently for different networks having different heterogeneity levels and fails to maintain the same stability period and lifetime as in previous heterogeneous WSNs. Some algorithms work efficiently in heterogeneous WSNs containing low energy difference between normal, advanced and super nodes and some algorithms work efficiently in networks containing high energy difference between normal, advanced and super nodes. So we interprets each algorithm in this paper, on basis of types of heterogeneous networks containing different heterogeneity level and parameters on basis of stability period, lifetime of network and packets sent to the BS.

\section{Heterogeneous WSN Model}
In this section, we assume $N$ number of nodes placed in a square region of dimension $M\times M$. Heterogeneous WSNs contain two, three or multi types of nodes with respect to their energy levels and are termed as two, three and multi level heterogeneous WSNs respectively.

\subsection{Two Level Heterogeneous WSNs Model}
Two level heterogeneous WSNs contain two energy level of nodes, normal and advanced nodes. Where, $E_{o}$ is the energy level of normal node and $E_{o}(1+a)$ is the energy level of advanced nodes containing $a$ times more energy as compared to normal nodes. If $N$ is the total number of nodes then $Nm$ is the number of advanced nodes where $m$ refers to the fraction of advanced nodes and $N(1-m)$ is the number of normal nodes. The total initial energy of the network is the sum of energies of normal and advanced nodes.

\begin{eqnarray}
\begin{split}
E_{total} = N(1-m)E_{o} + Nm(1+a)E_{o}\\ =NE_{o}(1-m+m+am)\\ =NE_{o}(1+am)
\end{split}
\end{eqnarray}

The two level heterogeneous WSNs contain $am$ times more energy as compared to homogeneous WSNs.

\subsection{Three Level Heterogeneous WSN Model}

Three level heterogeneous WSNs contain three different energy levels of nodes i.e normal, advanced and super nodes. Normal nodes contain energy of $E_{o}$, the advanced nodes of fraction m are having $a$ times extra energy than normal nodes equal to $E_{o}(1+a)$ whereas, super nodes of fraction $m_{o}$ are having a factor of $b$ times more energy than normal nodes so their energy is equal to $E_{o}(1+b)$. As $N$ is the total number of nodes in the network, then $Nmm_{o}$ is total number of super nodes and $Nm(1-m_{o})$ is total number of advanced nodes. The total initial energy of three level heterogeneous WSN is therefore given by:

\begin{eqnarray}
E_{total} = N(1-m)E_{o}+Nm(1-m_{o})(1+a)E_{o}+ Nm_{o}E_{o}(1+b)
\end{eqnarray}

\begin{eqnarray}
E_{total} = NE_{o}(1+m(a+m_{o}b))
\end{eqnarray}

The three level heterogeneous WSNs contain $(a+m_{o}b)$ times more energy as compared to homogeneous WSNs.

\subsection{Multilevel Heterogeneous WSN Model}
Multi level heterogeneous WSN is a network that contains nodes of multiple energy levels. The initial energy of nodes is distributed over the close set $[E_{o},E_{o}(1+a_{max})]$, where $E_{o}$ is the lower bound and $a_{max}$ is the value of maximal energy. Initially, node $S_{i}$  is equipped with initial energy of $E_{o}(1+a_{i})$, which is $a_{i}$ times more energy than the lower bound $E_{o}$. The total initial energy of multi-level heterogeneous networks is given by:

\begin{eqnarray}
E_{total}= \sum_{i=1}^{N}E_{o}(1+a_{i})= E_{o}(N+\sum_{i=1}^{N}a_{i})
\end{eqnarray}

CH nodes consume more energy as compared to member nodes so after some rounds energy level of all the nodes becomes different as compared to each other. Therefore, heterogeneity is introduced in homogeneous WSNs and the networks that contain heterogeneity are more important than homogeneous networks.

\section{Radio Dissipation Model}
The radio energy model describes that l bit message is transmitted over a distance $d$ as in [5.6], energy expended is then given by:

\begin{eqnarray}
E_{Tx}(l,d)=
\begin{cases}
l E_{elec}+l\varepsilon_{fs}d^{2},&  d<d_{o} \\
l E_{elec}+l \varepsilon_{mp}d^{4},& d\geq d_{o} \\
\end{cases}
\end{eqnarray}

Where, $E_{elec}$ is the energy dissipated per bit to run the transmitter or the receiver circuit. $d$ is the distance between sender and receiver. If this distance is less than threshold, free space$(fs)$ model is used else multi path$(mp)$ model is used. Now, total energy dissipated in the network during a round is given by [5,6]:
\small
\begin{eqnarray}
E_{round}= L(2NE_{elec}+NE_{DA}+k\varepsilon_{mp}d_{to BS}^{4}+N\varepsilon_{fs}d_{to CH}^{2})
\end{eqnarray}
\normalsize

Where,
K= number of clusters\\
$E_{DA}$= Data aggregation cost expended in CH\\
$d_{to}BS$= Average distance between the CH and BS\\
$d_{to}CH$= Average distance between the cluster members and the CH\\

\begin{eqnarray}
d_{to CH}= \frac{M}{\sqrt{2 \pi k}}, d_{to BS}= 0.765\frac{M}{2}
\end{eqnarray}

\begin{eqnarray}
k_{opt}=\frac{\sqrt{N}}{\sqrt{2\pi}}\sqrt{\frac{\varepsilon_{fs}}{\varepsilon_{mp}}}\frac{M}{d_{toBS}^{2}}
\end{eqnarray}

\section{Overview of Distributed Heterogenous Protocols}

\subsection{DEEC}
DEEC is designed to deal with nodes of heterogeneous WSNs. For CH selection, DEEC uses initial and residual energy level of nodes. Let $n_{i}$ denote the number of rounds to be a CH for node $s_{i}$. $p_{opt}N$ is the optimum number of CHs in our network during each round. CH selection criteria in DEEC is based on energy level of nodes. As in homogenous network, when nodes have same amount of energy during each epoch then choosing $p_{i}=p_{opt}$ assures that $p_{opt}N$ CHs during each round. In WSNs, nodes with high energy are more probable to become CH than nodes with low energy but the net value of CHs during each round is equal to $p_{opt}N$. $p_{i}$ is the probability for each node $s_{i}$ to become CH, so, node with high energy has larger value of $p_{i}$ as compared to the $p_{opt}$. $\bar{E}(r)$ denotes average energy of network during round $r$ which can be given as in [10]:

\begin{eqnarray}
\bar{E}(r)= \frac{1}{N}\sum_{i=1}^{N}E_{i}(r)
\end{eqnarray}

Probability for CH selection in DEEC is given as in [10]:

\begin{eqnarray}
p_{i}= p_{opt}[1-\frac{\bar{E}(r)-E_{i}(r)}{\bar{E}(r)}] = p_{opt}\frac{E_{i}(r)}{\bar{E}(r)}
\end{eqnarray}
In DEEC the average total number of CH during each round is given as in [10]:

\begin{eqnarray}
\begin{split}
\sum_{i=1}^{N}p_{i}= \sum_{i=1}^{N}p_{opt}\frac{E_{i}(r)}{\bar{E}(r)} = p_{opt}\sum_{i=1}^{N}\frac{E_{i}(r)}{\bar{E}(r)}= Np_{opt}
\end{split}
\end{eqnarray}

$p_{i}$ is probability of each node to become CH in a round. Where $G$ is the set of nodes eligible to become CH at round r. If node becomes CH in recent rounds then it belongs to G. During each round each node chooses a random number between 0 and 1. If number is less than threshold as defined in equation 12 as in [10], it is eligible to become a CH else not.

\begin{eqnarray}
T(s_{i})=
\begin{cases}
\frac{p_{i}}{1-p_{i}(rmod\frac{1}{P_{i}})} & if\; s_{i}\epsilon G \\
0 & otherwise
\end{cases}
\end{eqnarray}

As $p_{opt}$ is reference value of average probability $p_{i}$. In homogenous networks, all nodes have same initial energy so they use $p_{opt}$ to be the reference energy for probability $p_{i}$. However in heterogeneous networks, the value of  $p_{opt}$ is different according to the initial energy of the node.
In two level heterogenous network the value of $p_{opt}$ is given by as in [10]:

\begin{eqnarray}
p_{adv}= \frac{p_{opt}}{1+am} , p_{nrm}= \frac{p_{opt}(1+a)}{(1+am)}
\end{eqnarray}

Then use the above $p_{adv}$ and $p_{nrm}$ instead of $p_{opt}$ in equation $10$ for two level heterogeneous network as supposed in [10]:

\begin{eqnarray}
p_{i}=
\begin{cases}
\frac{p_{opt}E_{i}(r)}{(1+am)\bar{E}(r)}   &   if \;s_{i}\; is\; the\; normal\; node\\
\frac{p_{opt}(1+a)E_{i}(r)}{(1+am)\bar{E}(r)}   &   if \;s_{i}\; is\; the\; advanced \;node\\
\end{cases}
\end{eqnarray}

Above model can also be extended to multi level heterogenous network given below as in [10]:

\begin{eqnarray}
p_{multi}= \frac{p_{opt}N(1+a_{i})}{(N+\sum_{i=1}^{N}a_{i})}
\end{eqnarray}

Above $p_{multi}$  in equation 10 instead of $p_{opt}$ to get $p_{i}$ for heterogeneous node. $p_{i}$ for the multilevel heterogeneous network is given by as in [10]:

\begin{eqnarray}
  p_{i}= \frac{p_{opt}N(1+a)E_{i}(r)}{(N+\sum_{i=1}^{N}a_{i})\bar{E}(r)}
\end{eqnarray}

In DEEC we estimate average energy $E(r)$ of the network for any round $r$ as in [10]:

\begin{eqnarray}
\bar{E}(r)= \frac{1}{N}E_{total}(1-\frac{r}{R})
\end{eqnarray}

R denotes total rounds of network lifetime and is estimated as follows:

\begin{eqnarray}
R= \frac{E_{total}}{E_{round}}
\end{eqnarray}

$E_{total}$ is total energy of the network where $E_{round}$ is energy expenditure during each round.
\subsection{DDEEC}
DDEEC uses same method for estimation of average energy in the network and CH selection algorithm based on residual energy as implemented in DEEC.
Difference between DDEEC and DEEC is centered in expression that defines probability for normal and advanced nodes to be a CH [11] as given in equation 14.\\
We find that nodes with more residual energy at round $r$ are more probable to become CH, so, in this way nodes having higher energy values or advanced nodes will become CH more often as compared to the nodes with lower energy or normal nodes. A point comes in a network where advanced nodes having same residual energy like normal nodes. Although, after this point DEEC continues to punish the advanced nodes so this is not optimal way for energy distribution because by doing so, advanced nodes are continuously a CH and they die more quickly than normal nodes. To avoid this unbalanced case, DDEEC makes some changes in equation 14 to save advanced nodes from being punished over and again. DEEC introduces threshold residual energy as in [11] and given below:

\begin{eqnarray}
Th_{REV}= E_{o}(1+\frac{a E_{disNN}}{E_{disNN}-E_{disAN}})
\end{eqnarray}

When energy level of advanced and normal nodes falls down to the limit of threshold residual energy then both type of nodes use same probability to become  cluster head. Therefore, CH selection is balanced and more efficient. Threshold residual energy $Th$ is given as in [11] and given below:

\begin{eqnarray}
Th_{REV}\simeq (7/10)E_{o}
\end{eqnarray}

Average probability $p_{i}$ for CH selection used in DDEEC is as follows as in [11]:
\small
\begin{eqnarray}
p_{i}=
\begin{cases}
 \frac{p_{opt}E_{i}(r)}{(1+am)\bar{E}(r)} & for  \;   Nml  \;   nodes,  \;E_{i}(r)>Th_{REV} \\
 \frac{(1+a)p_{opt}E_{i}(r)}{(1+am)\bar{E}(r)} & for  \;  Adv   \;  nodes, \;E_{i}(r)>Th_{REV}\\
 c\frac{(1+a)p_{opt}E_{i}(r)}{(1+am)\tilde{}\bar{E}(r)} & for\; Adv   , \;  Nml  \;   nodes,\; E_{i}(r)\leq Th_{REV}\\
 \end{cases}
\end{eqnarray}
\normalsize

\subsection{EDEEC}
EDEEC uses concept of three level heterogeneous network as described above. It contains three types of nodes normal, advanced and super nodes based on initial energy. $p_{i}$ is probability used for CH selection and $p_{opt}$ is reference for $p_{i}$. EDEEC uses different $p_{opt}$ values for normal, advanced and super nodes, so, value of $p_{i}$ in EDEEC is as follows as in [12]:

\begin{eqnarray}
p_{i}=
\begin{cases}
\frac{p_{opt}E_{i}(r)}{(1+m(a+m_{o}b))\bar{E}(r)}   & if \; s_{i} \;is\; the\; normal \;node\\
\frac{p_{opt}(1+a)E_{i}(r)}{(1+m(a+m_{o}b))\bar{E}(r)}  & if \;s_{i}\; is \;the \;advanced \;node\\
\frac{p_{opt}(1+b)E_{i}(r)}{(1+m(a+m_{o}b))\bar{E}(r)}  & if \;s_{i} \; is \;the\; super \;node\\
\end{cases}
\end{eqnarray}

Threshold for CH selection for all three types of node is as follows as in [12]:

\begin{eqnarray}
T(s_{i})=
\begin{cases}
\frac{p_{i}}{1-p_{i(rmod\frac{1}{p_{i}})}}  &   if p_{i}\epsilon G'\\
\frac{p_{i}}{1-p_{i}(rmod\frac{1}{p_{i}})}  &   if p_{i}\epsilon G''\\
\frac{p_{i}}{1-p_{i}(rmod\frac{1}{p_{i}})}  &   if p_{i}\epsilon G'''\\
0   &   otherwise
\end{cases}
\end{eqnarray}
\subsection{TDEEC}
TDEEC uses same mechanism for CH selection and average energy estimation as proposed in DEEC. At each round, nodes decide whether to become a CH or not by choosing a random number between 0 and 1. If number is less than threshold $T_{s}$ as shown in equation 24 then nodes decide to become a CH for the given round.
In TDEEC, threshold value is adjusted and based upon that value a node decides whether to become a CH or not by introducing residual energy and average energy of that round with respect to optimum number of CHs [13].
Threshold value proposed by TDEEC is given as follows as in [13]:
\small
\begin{eqnarray}
T(s)= \{\frac{p}{1-p(rmod\frac{1}{p})}*\frac{residual \;energy \;of \;a \;node*k_{opt}}{average \;energy \;of \;the \; network}
\end{eqnarray}
\normalsize

\section{Performance Criteria}
Performance parameters used for evaluation of clustering protocols for heterogeneous WSNs are lifetime of heterogeneous WSNs, number of nodes alive during rounds and data packets sent to BS.\\

$Lifetime$ is a parameter which shows that node of each type has not yet consumed all of its energy.\\

$Number \;of\; nodes\; alive$ is a parameter that describes number of alive nodes during each round.\\

$Data\; packets\; sent \;to \;the \;BS$ is the measure that how many packets are received by BS for each round.\\

These parameters depict stability period, instability period, energy consumption, data sent to the BS, and data received by BS and lifetime of WSNs. Stability period is period from start of network until the death of first node whereas, instability period is period from the death of first node until last one.

\begin{table}[!h]
\caption{value of parameters}
\begin{center}
    \begin{tabular}{ | p{2.5cm} | p{2.5cm} |}
    \hline
    Parameters &	Values\\ \hline
    Network field & 100 m,100 m\\ \hline
    Number of nodes	& 100\\ \hline
    $E_{o}$(initial energy of normal nodes) & 0.5J\\ \hline
    Message size &	4000 bits \\ \hline
    $E_{elec}$ & 50nJ/bit\\ \hline
    $E_{fs}$ & 10nJ/bit/m2\\ \hline
    $E_{amp}$ & 0.0013pJ/bit/m4\\ \hline
    $EDA$ & 5nJ/bit/signal\\ \hline
    $d_{o}$(threshold distance) & 70m\\ \hline
    $P_{opt}$ & 0.1 \\ \hline
    \end{tabular}
\end{center}
\end{table}
\section{Simulations And Discussions}
In this section, we simulate different clustering protocols in heterogeneous WSN using MATLAB and for simulations we use 100 nodes randomly placed in a field of dimension $100m\times 100m$. For simplicity, we consider all nodes are either fixed or micro-mobile as supposed in [14] and ignore energy loss due to signal collision and interference between signals of different nodes that are due to dynamic random channel conditions. In this scenario, we are considering that, BS is placed at center of the network field. We simulate DEEC, DDEEC, EDEEC and TDEEC for three-level and multi-level heterogeneous WSNs. Scenarios describe values for  number of nodes dead in first, tenth and last rounds as well as values for the packets sent to BS by CH at different values of parameters $m$, $m_{o}$, $a$ and $b$. These values are examined for DEEC, DDEEC, EDEEC and TDEEC.\\
\indent In heterogeneous WSN, we use radio parameters mentioned in Table 1 for different protocols deployed in WSN and estimate the performance for three level heterogeneous WSNs. Parameter $m$ refers to fraction of advanced nodes containing extra amount of energy $a$ in network whereas, $m_{o}$ is a factor that refers to fraction of super nodes containing extra amount of energy $b$ in the network.

\begin{figure}[h!]
\center
\includegraphics[height=5cm, width=7cm]{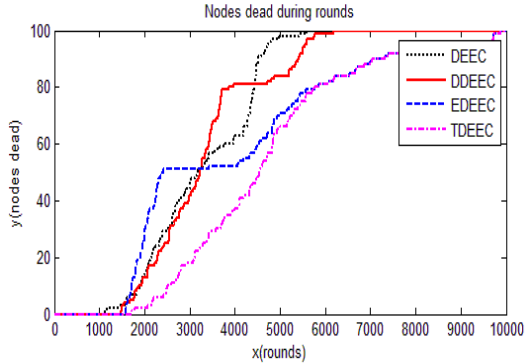}
\caption{Nodes dead during rounds}
\end{figure}

\begin{figure}[h!]
\center
\includegraphics[height=5.2cm, width=7cm]{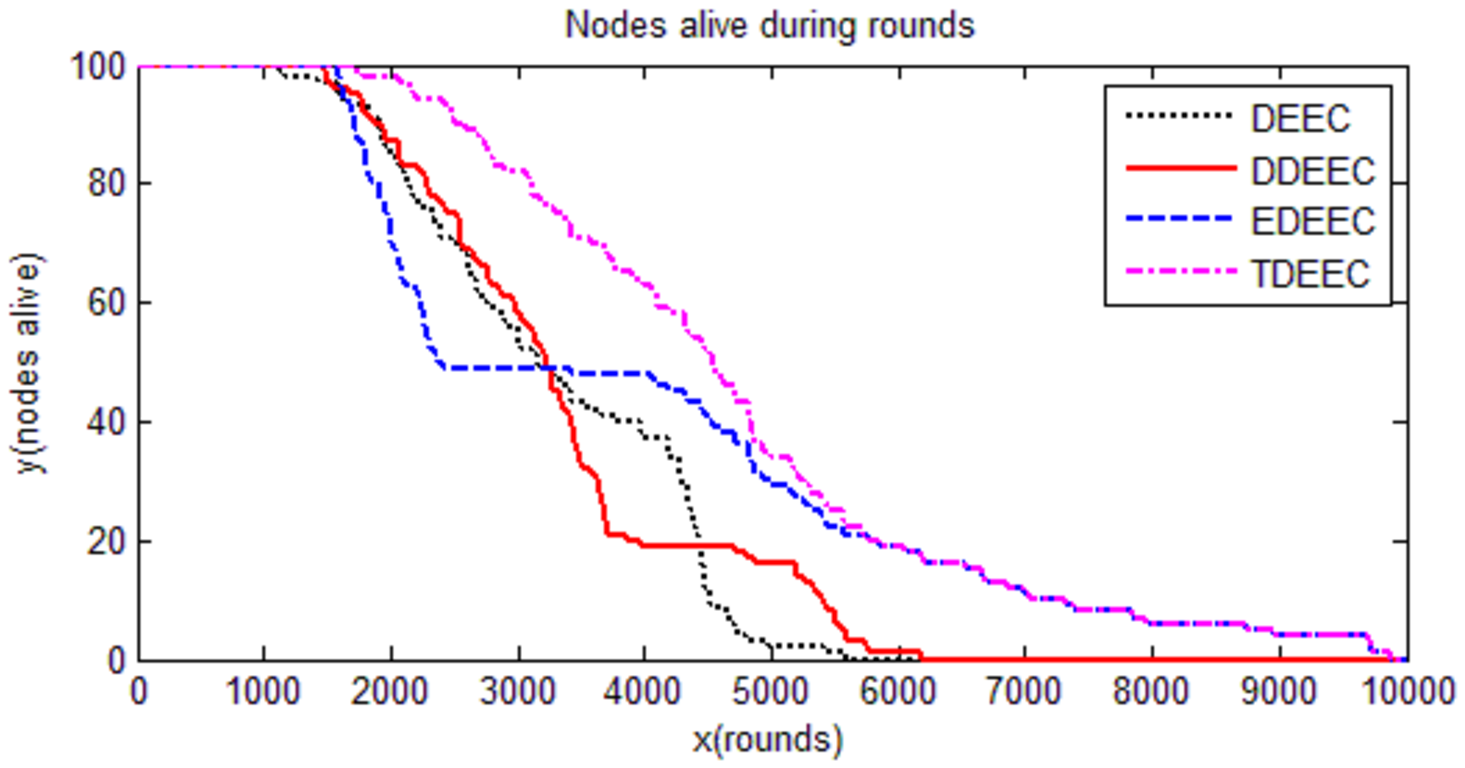}
\caption{Nodes alive during rounds}
\end{figure}

\begin{figure}[h!]
\center
\includegraphics[height=5.2cm, width=7cm]{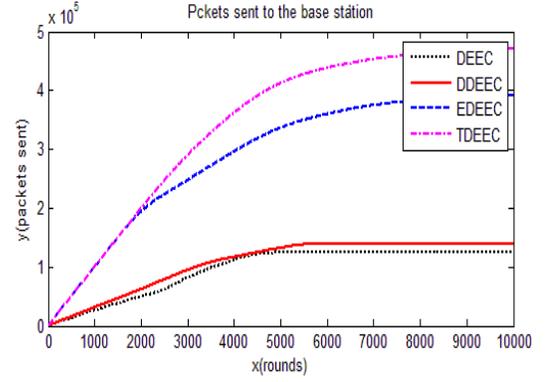}
\caption{Packets sent to the BS}
\end{figure}

For the case of a network containing $m=0.5$ fraction of advanced nodes having $a=1.5$ times more energy and $m_{o}$=0.4 fraction of super nodes containing $b=3$ times more energy than normal nodes. From Fig. 1 and 2, we examine that first node for DEEC, DDEEC, EDEEC and TDEEC dies at 1117, 1470, 1583 and 1719 rounds respectively. Tenth node dies at 1909, 1863, 1726 and 1297 rounds respectively. All nodes are dead at 5588, 6180, 9873 and 9873 rounds respectively. It is obvious from the results of all protocols that in terms of stability period, TDEEC performs best of all, EDEEC performs better than DEEC and DDEEC but has less performance than TDEEC. DDEEC only performs well as compared to DEEC and DEEC has least performance than all the protocols. Stability period of DEEC and DDEEC is lower than EDEEC and TDEEC because the probabilities in TDEEC and EDEEC are defined separately for normal, advanced and super nodes whereas, DEEC and DDEEC do not use different probabilities for normal, advanced and super nodes so their performance is lower than EDEEC and TDEEC. However, instability period of EDEEC and TDEEC is much larger than DEEC and DDEEC. The number of nodes alive in TDEEC is quite larger than EDEEC because in TDEEC the formula of threshold used by nodes for CH election is modified by including residual and average energy of that round. So nodes having high energy will become CHs. Similarly, by examining results of Fig. 3, packets sent to the BS by DEEC, DDEEC, EDEEC and TDEEC have their values at 125316, 139314, 391946 and 470248. Now we see that packets sent to BS for DEEC and DDEEC is almost same whereas, the packets sent to BS for EDEEC and TDEEC are almost the same because the probability equations for normal, advanced and super nodes is same in both of them. Now coming to the CHs, the packets sent to CHs increase during the start of the network and gradually decrease down towards the end due to the nodes dying simultaneously.\\

\begin{figure}[h!]
\center
\includegraphics[height=5.2cm, width=7cm]{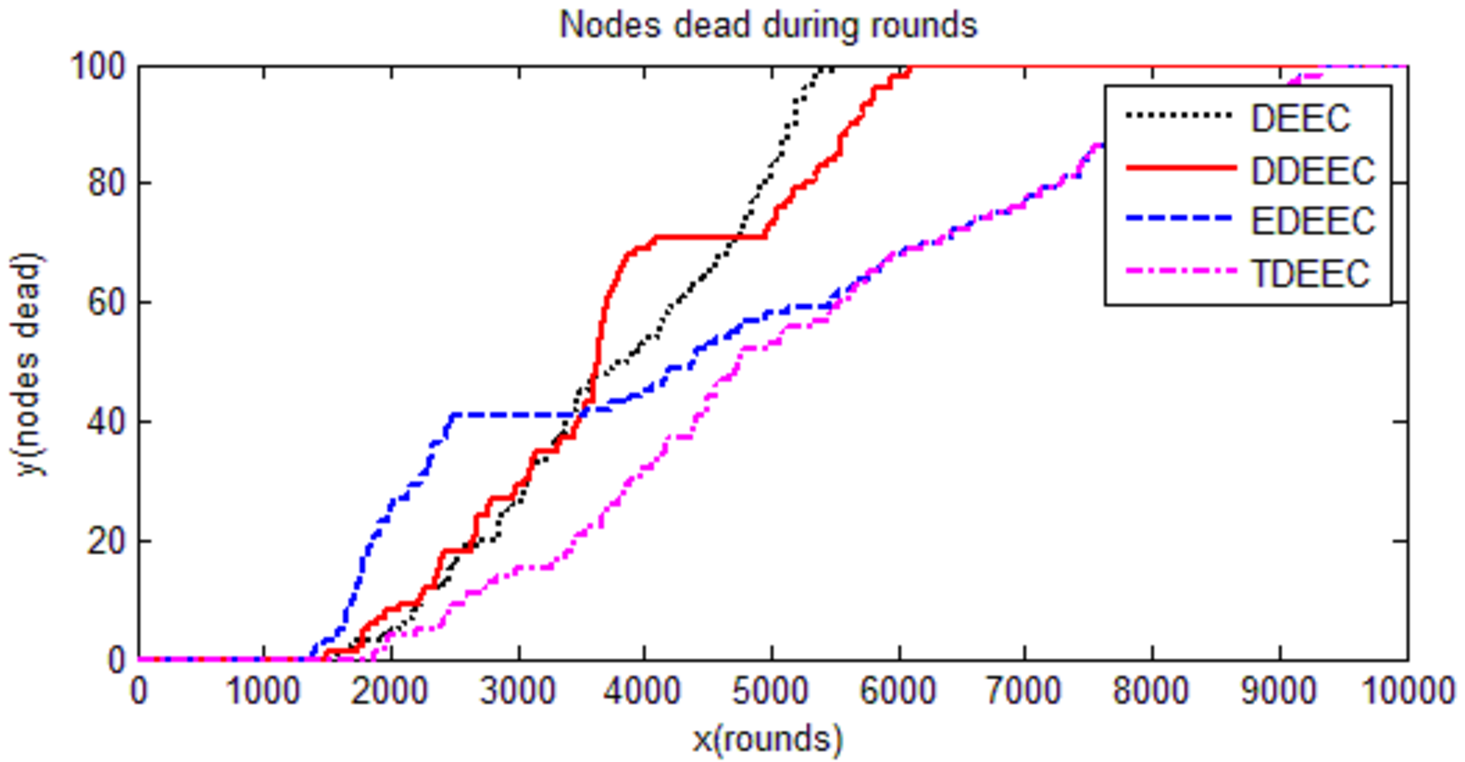}
\caption{Nodes dead during rounds}
\end{figure}
\begin{figure}[h!]
\center
\includegraphics[height=5.2cm, width=7cm]{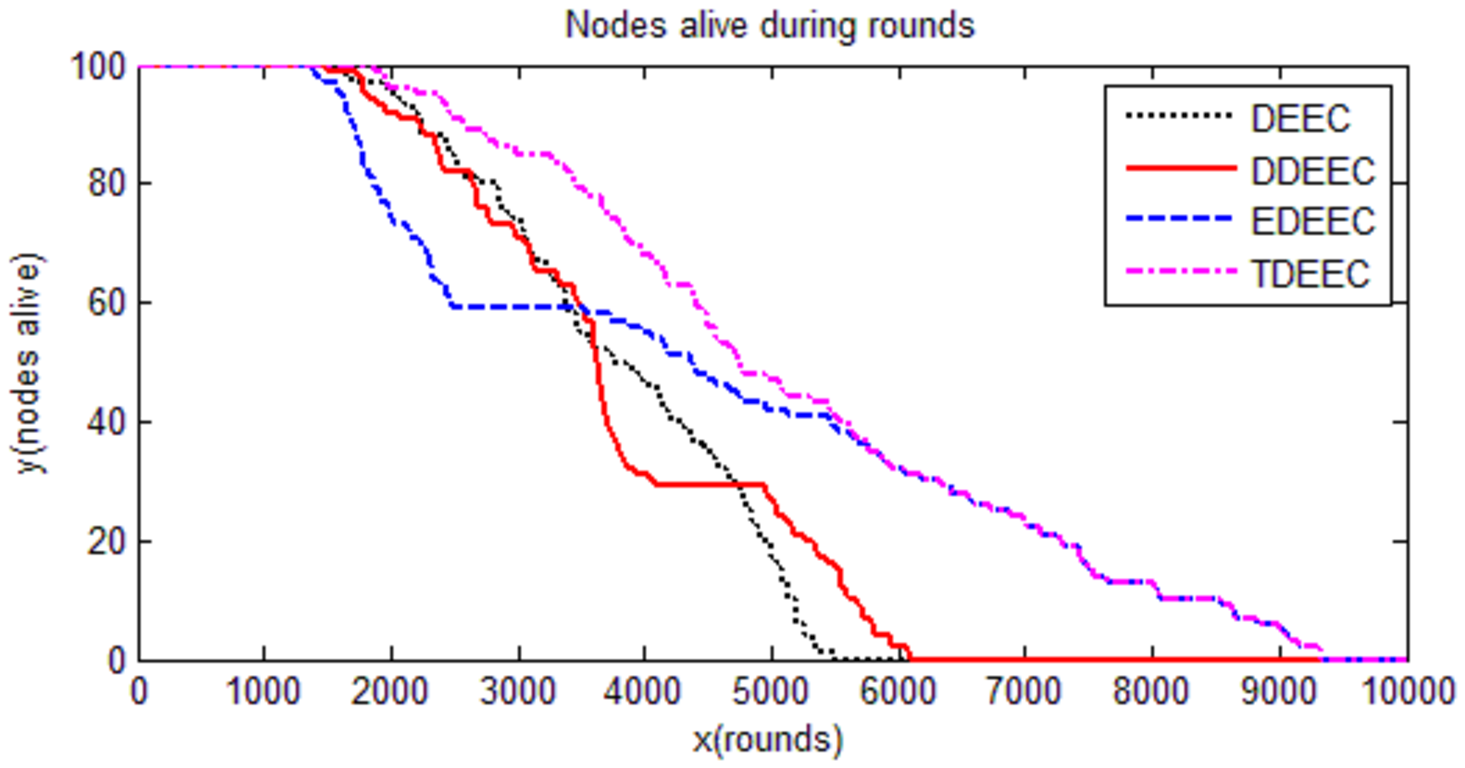}
\caption{Nodes alive during rounds}
\end{figure}
\begin{figure}[h!]
\center
\includegraphics[height=5.2cm, width=7cm]{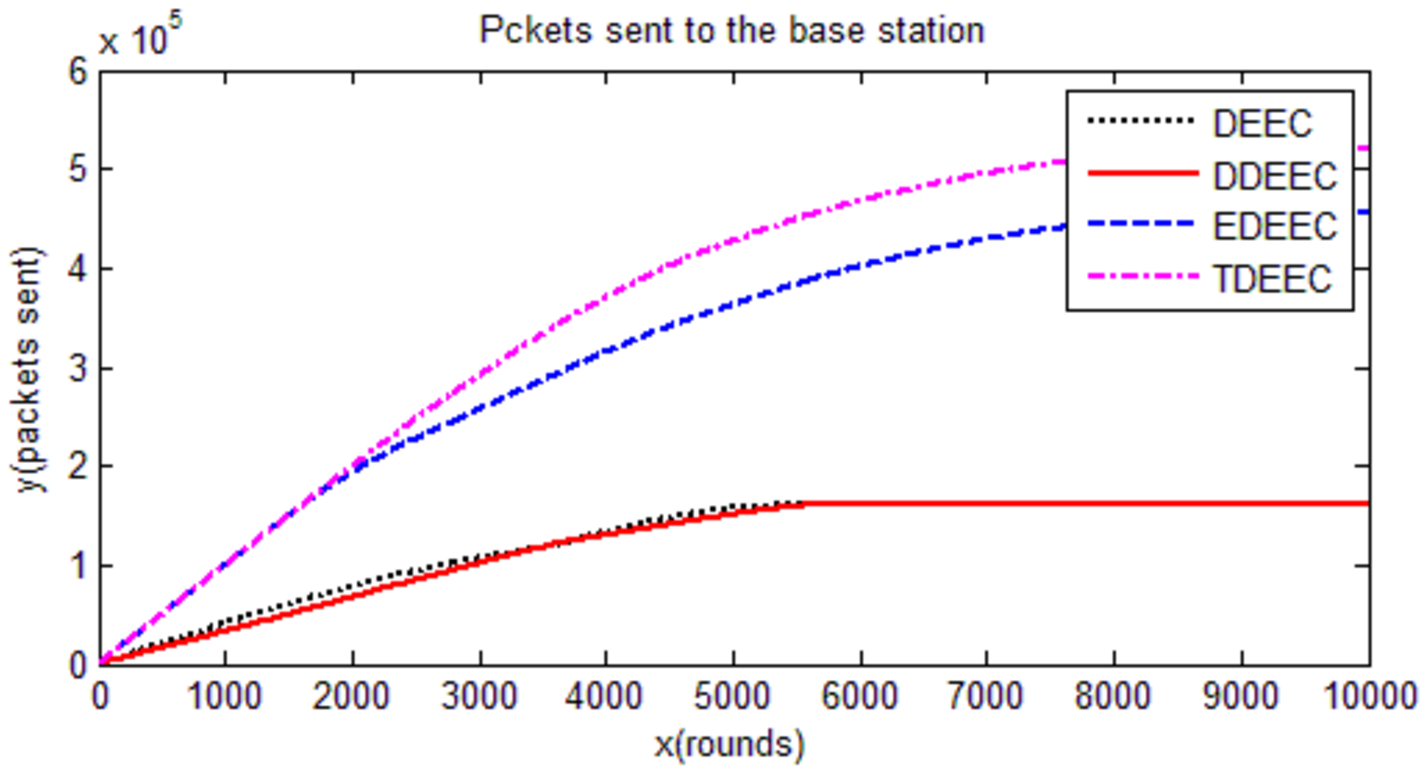}
\caption{Packets sent to the BS}
\end{figure}

\indent Now considering second case in which the parameters change to $a=1.3$, $b=2.5$, $m=0.4$ and $m_{o}=0.3$. Fig. 4 shows that first node for DEEC, DDEEC, EDEEC and TDEEC dies of each protocol at 1291, 1355, 1367 and 1694 rounds respectively. Tenth node dies at 1531, 1547, 1574 and 1946 rounds respectively. All nodes are dead at 4870, 4779, 7291, 7291 rounds. Graph for number of nodes alive in first, tenth and all rounds is exactly the flip to the graph for number of nodes dead and is shown in Fig. 5. Result of Fig. 6 shows that packets sent to BS by DEEC, DDEEC, EDEEC and TDEEC are 135650, 107891, 300735 and 365628 respectively. As we see that by decreasing the values of parameters, TDEEC still performs best among the four protocols. EDEEC performs better than TDEEC. DDEEC performs better than TDEEC and EDEEC whereas, DEEC performs worst.\\

\begin{figure}[h!]
\center
\includegraphics[height=5.2cm, width=7cm]{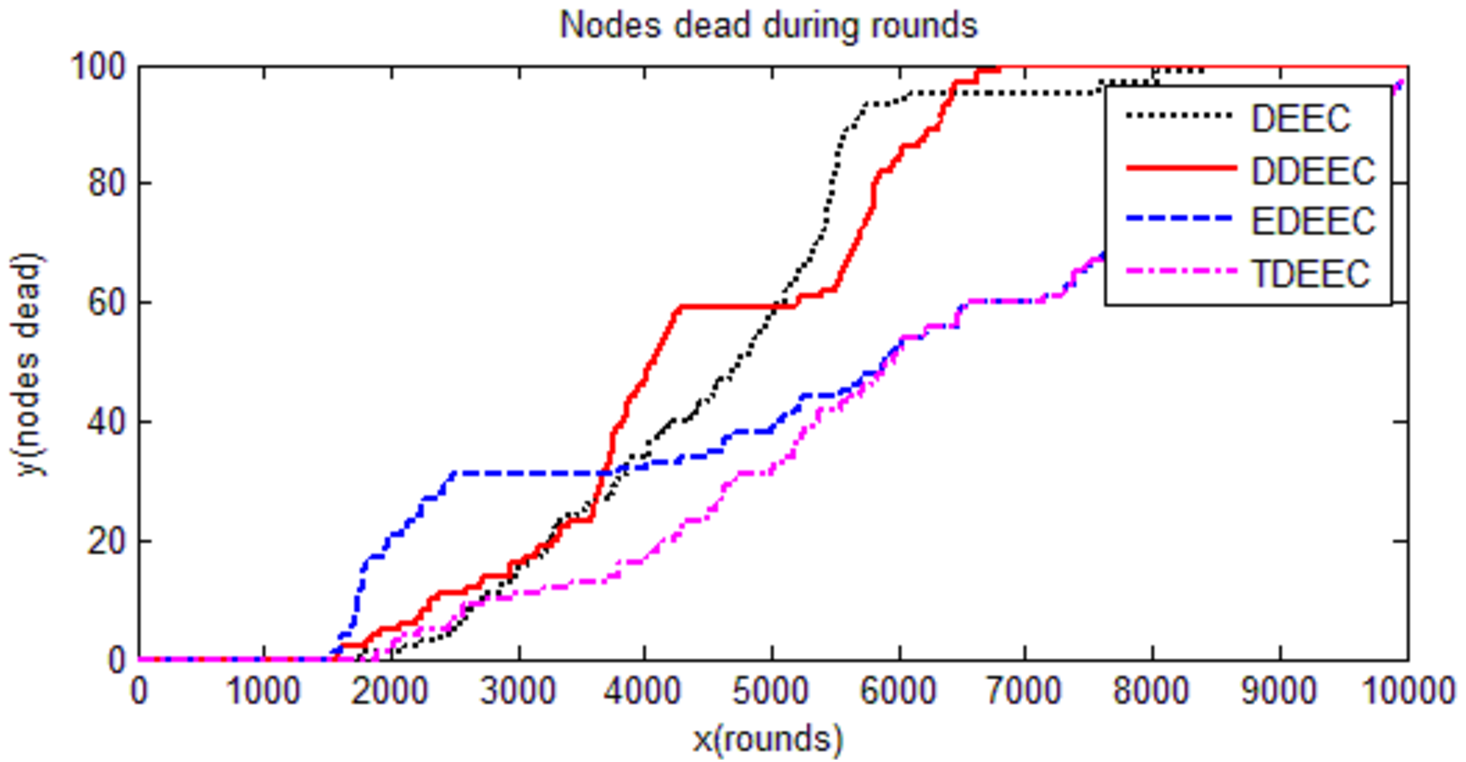}
\caption{Nodes dead during rounds}
\end{figure}
\begin{figure}[h!]
\center
\includegraphics[height=5.2cm, width=7cm]{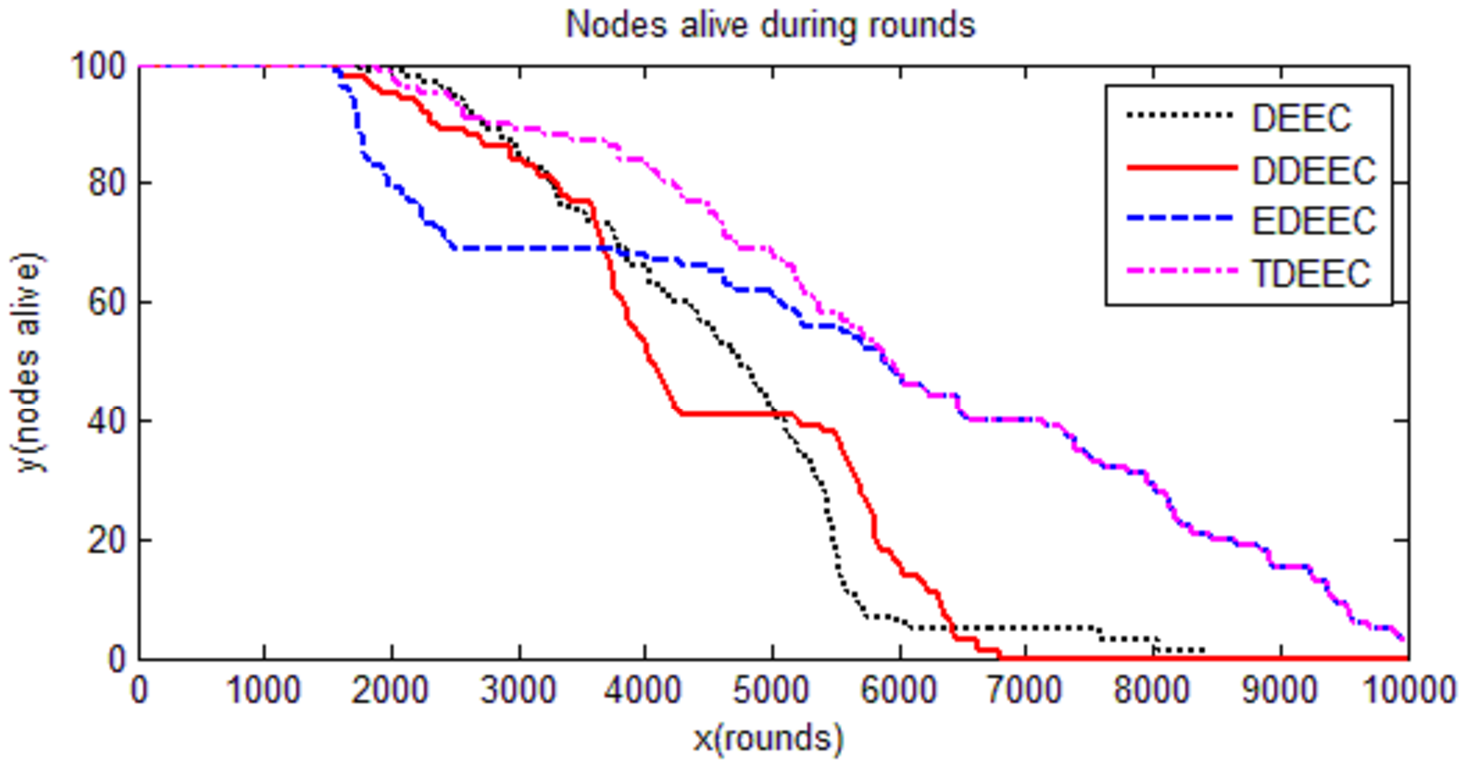}
\caption{Nodes alive during rounds}
\end{figure}
\begin{figure}[h!]
\center
\includegraphics[height=5.2cm, width=7cm]{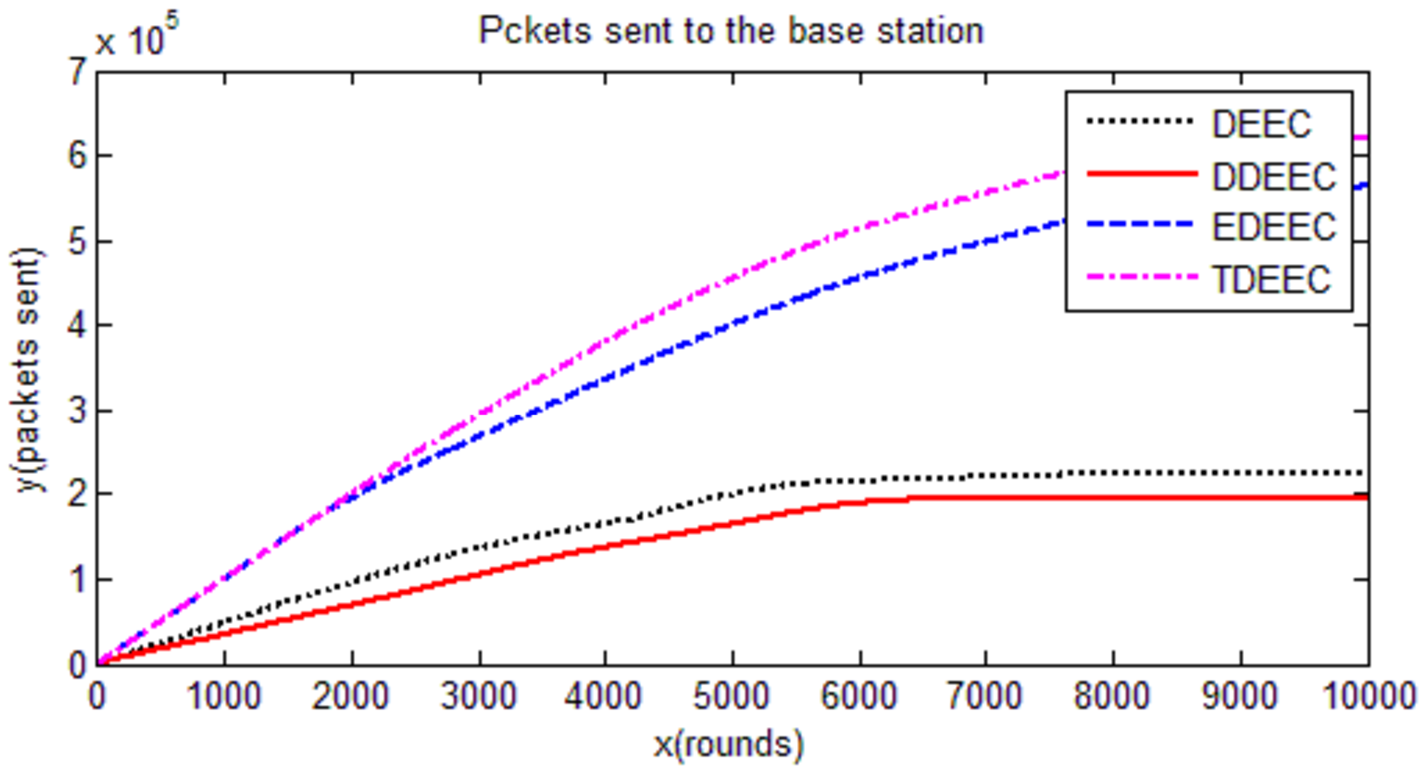}
\caption{Packets sent to the BS}
\end{figure}
\indent Now considering third case, parameter values further decrease to $a=1.2$, $b=2$, $m=0.3$, $m_{o}=0.2$ in which first node for DEEC, DDEEC, EDEEC and TDEEC dies at 963, 1158, 1309, and 1753 rounds respectively. Tenth node dies at 1290, 1573, 1556 and 2026 rounds respectively. All nodes are dead at 6533, 4386, 7467 and 7467 rounds respectively. Similarly, the packets to BS sent in DEEC, DDEEC, EDEEC and TDEEC are 132378, 91269, 259370 and 339406 respectively as shown in Fig. 7, 8 and 9.\\
\begin{figure}[h!]
\center
\includegraphics[height=5.2cm, width=7cm]{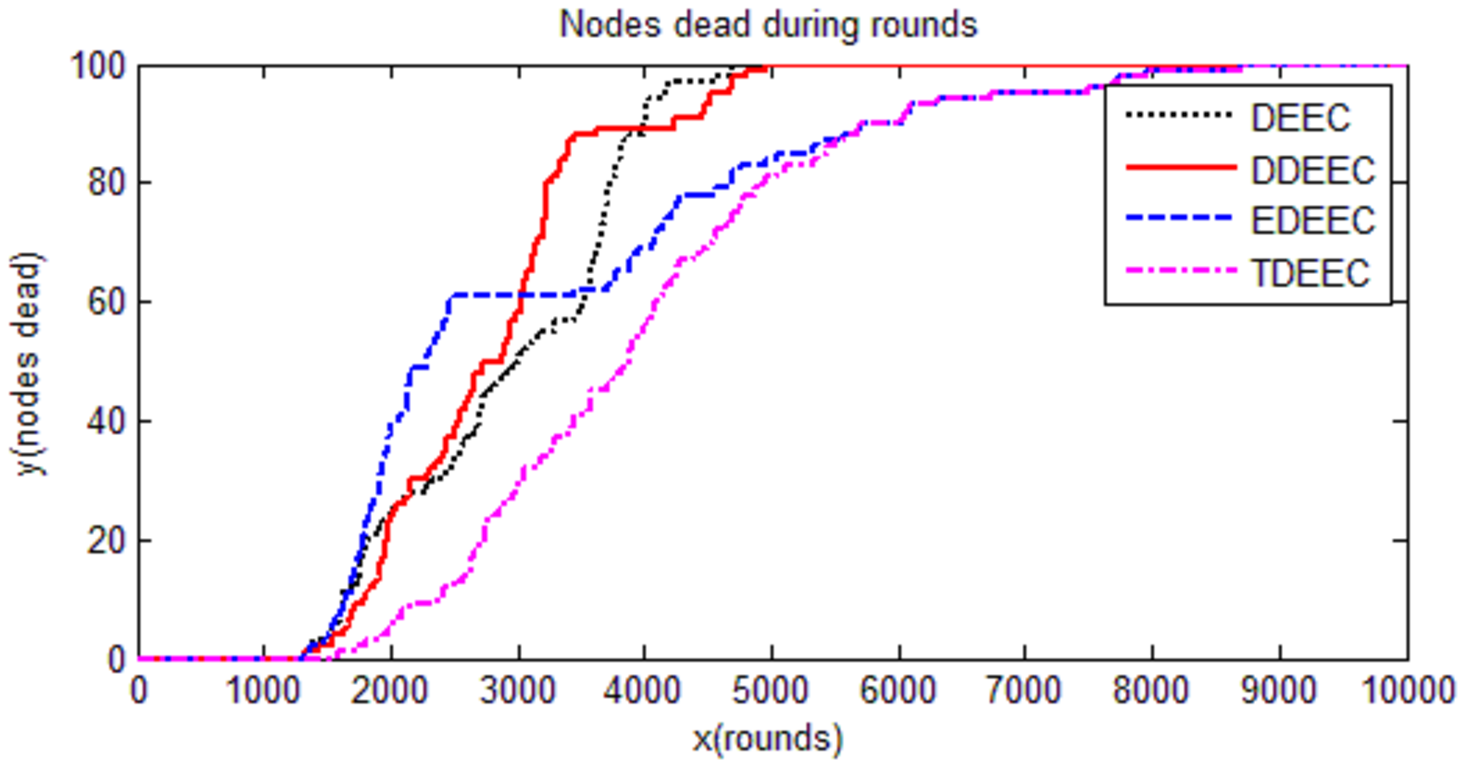}
\caption{Nodes dead during rounds}
\end{figure}
\begin{figure}[h!]
\center
\includegraphics[height=5.2cm, width=7cm]{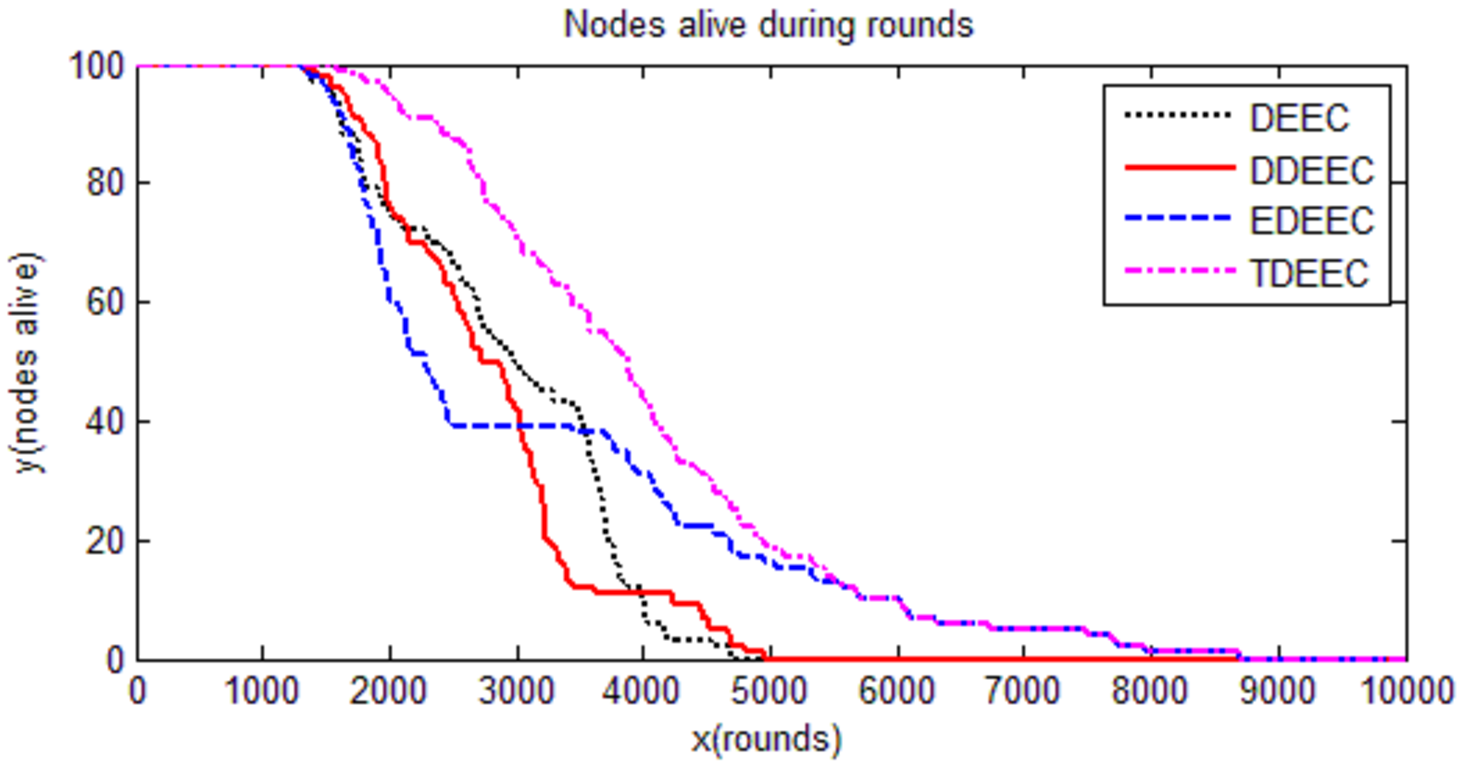}
\caption{Nodes alive during rounds}
\end{figure}
\begin{figure}[h!]
\center
\includegraphics[height=5.2cm, width=7cm]{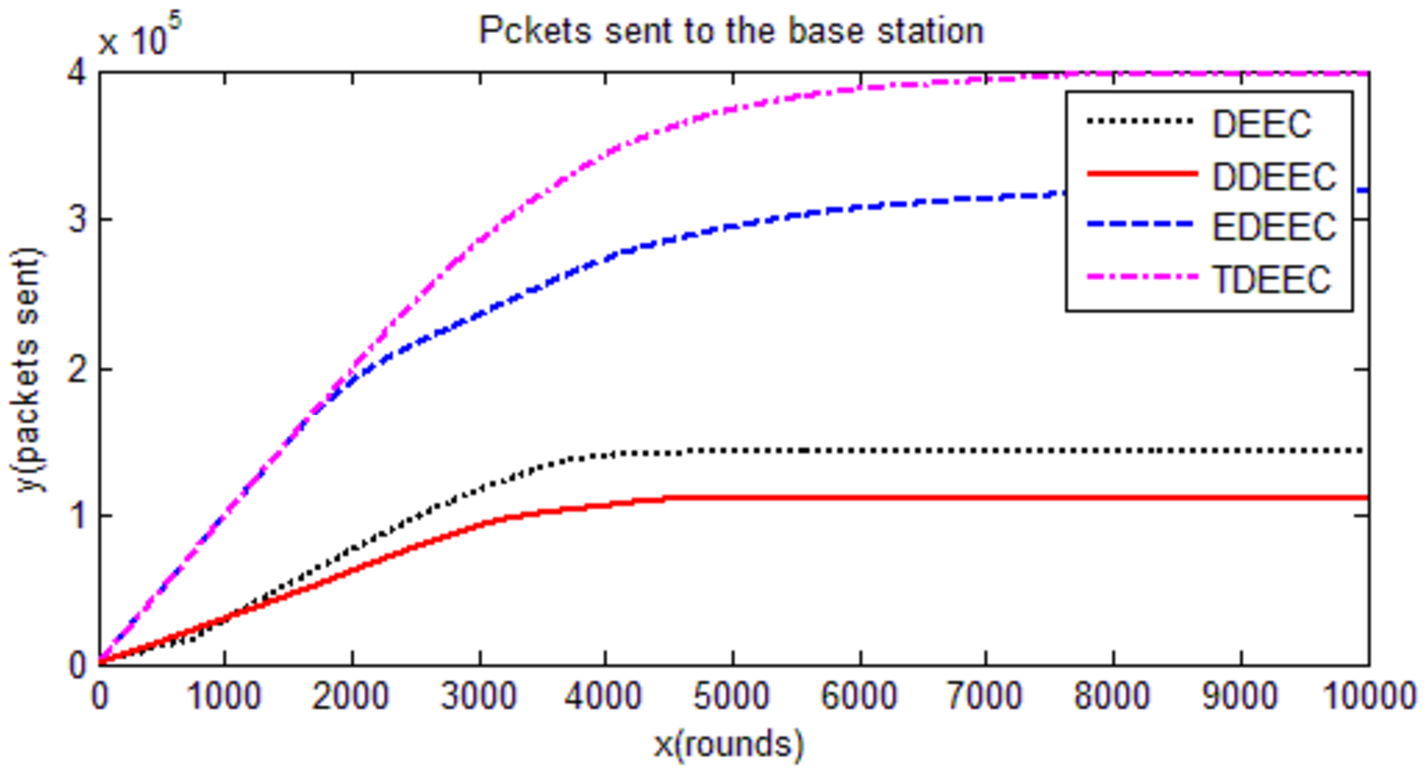}
\caption{Packets sent to the BS}
\end{figure}
\indent Now considering fourth case, parameters are increased to $a=1.6$, $b=3.2$, $m=0.6$, $m_{o}=0.5$. Results show that for DEEC, DDEEC, EDEEC and TDEEC first node dies at 1576, 1495, 1382 and 1863 round respectively. Tenth node dies at 2245, 2213, 1691 and 2574 round respectively and all nodes are dead at 5498, 6092, 9331 and 9331 round respectively. Packets sent to the BS in DEEC, DDEEC, EDEEC and TDEEC are 116181, 162506, 455423 and 521450 respectively as shown in Fig. 10, 11 and 12.\\

\begin{figure}[h!]
\center
\includegraphics[height=5.2cm, width=7cm]{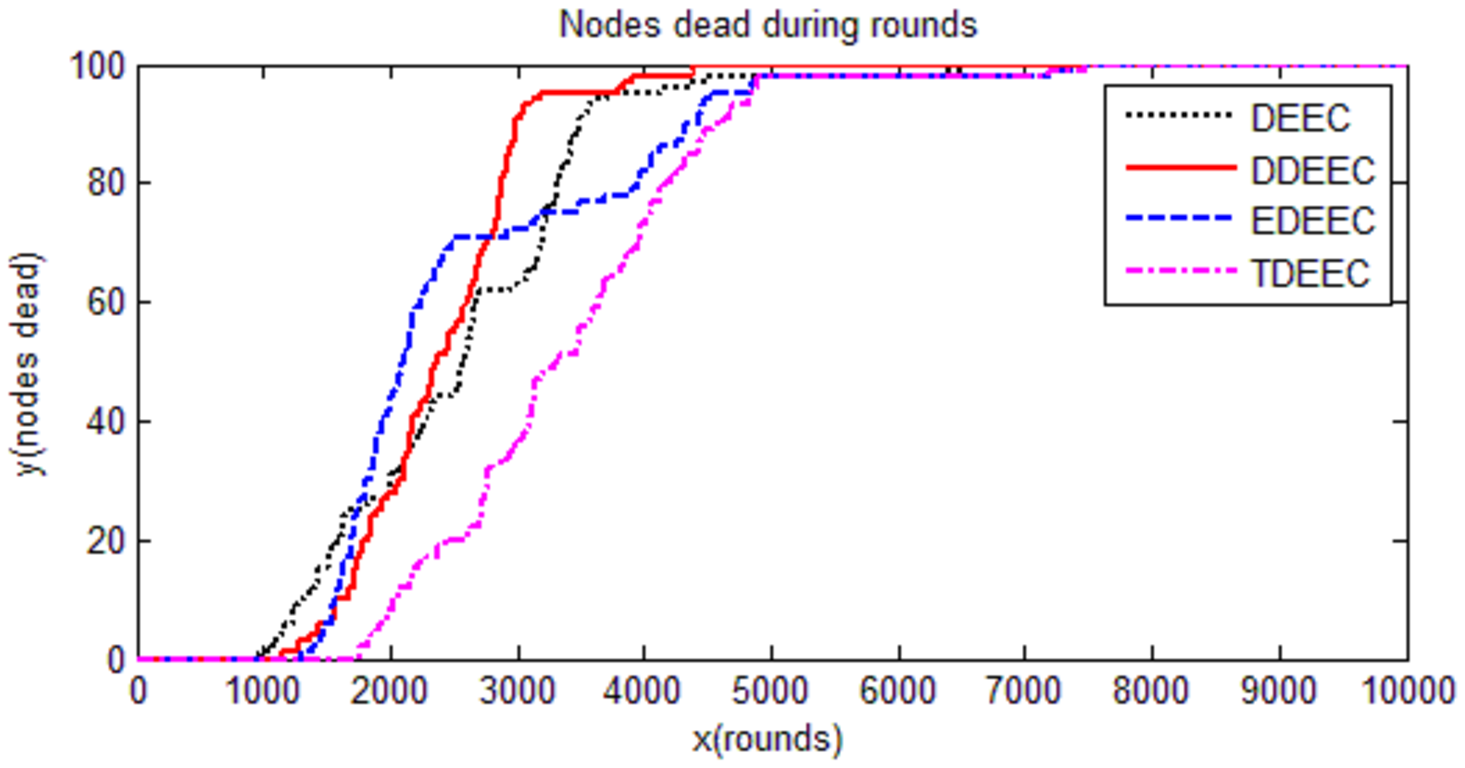}
\caption{Nodes dead during rounds}
\end{figure}
\begin{figure}[h!]
\center
\includegraphics[height=5.2cm, width=7cm]{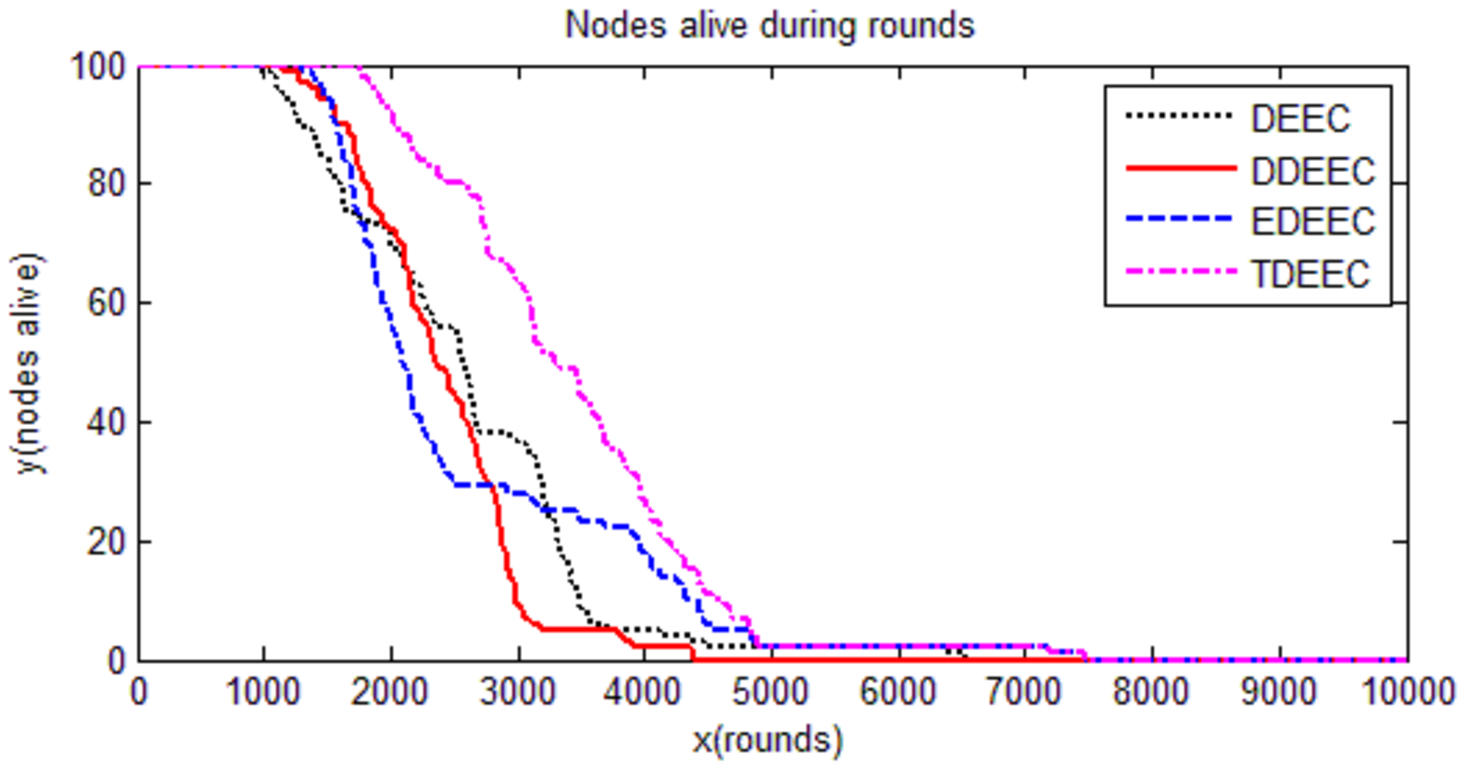}
\caption{Nodes alive during rounds}
\end{figure}
\begin{figure}[h!]
\center
\includegraphics[height=5.2cm, width=7cm]{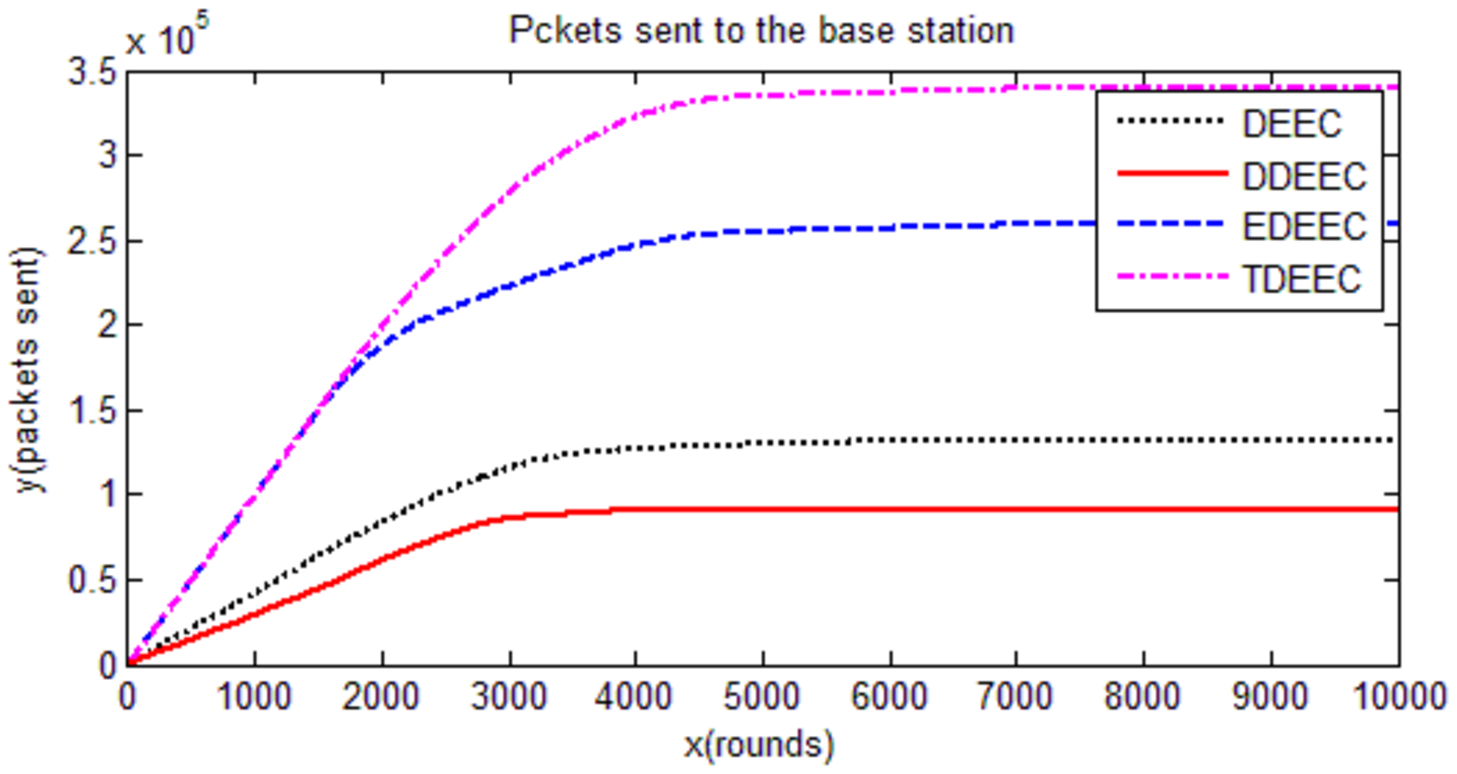}
\caption{Packets sent to the BS}
\end{figure}
\indent Now considering the fifth case and further more increasing the parameters to $a=1.7$, $b=3.4$, $m=0.7$, $m_{o}=0.6$ it is observed that for DEEC, DDEEC, EDEEC and TDEEC first node dies at 1763, 1584, 1551, 1897 rounds respectively. Tenth node dies at 2711, 2308, 1735, 2725 rounds respectively. All nodes dead for DEEC and DDEEC are  8414, 6786 rounds and for EDEEC ,TDEEC still some nodes are alive after 10000 rounds. Packets sent to the BS in DEEC, DDEEC, EDEEC and TDEEC are 224095, 193931, 562819, 620606 respectively as shown in Fig. 13, 14 and 15.

\begin{figure}[h!]
\center
\includegraphics[height=5.2cm, width=7cm]{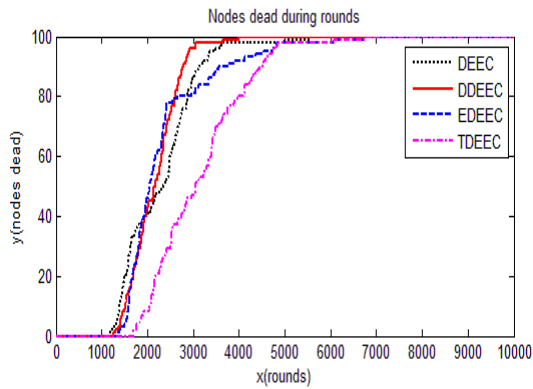}
\caption{Nodes dead during rounds}
\end{figure}

\begin{figure}[h!]
\center
\includegraphics[height=5.2cm, width=7cm]{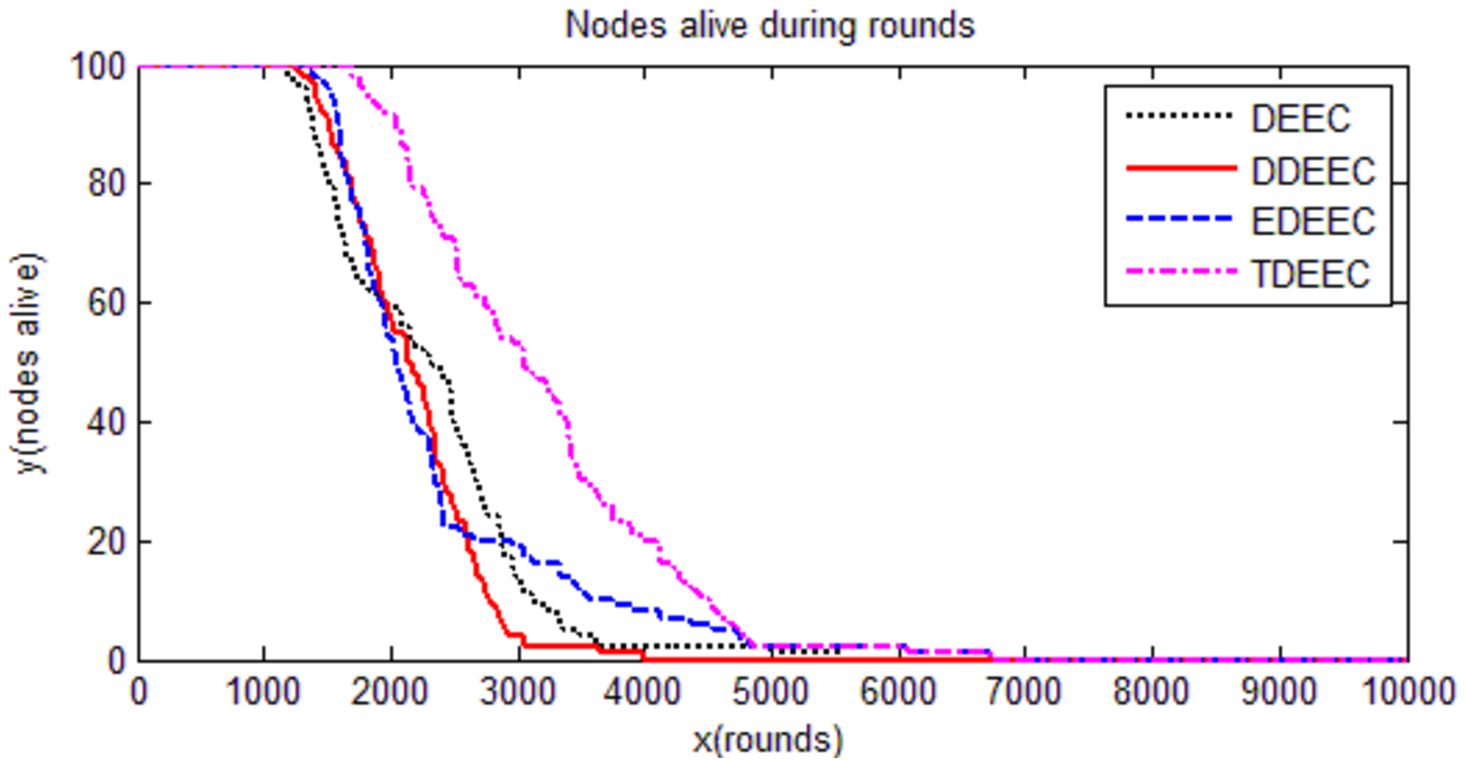}
\caption{Nodes alive during rounds}
\end{figure}
\begin{figure}[h!]
\center
\includegraphics[height=5.2cm, width=7cm]{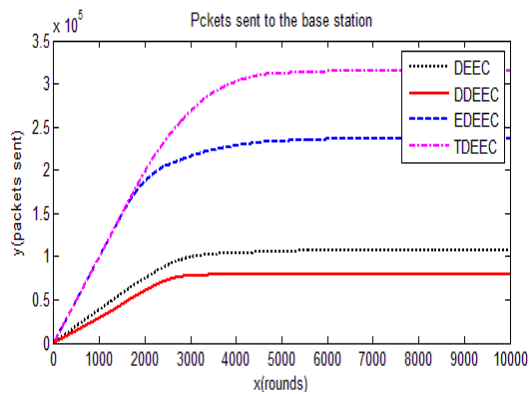}
\caption{Packets sent to the BS}
\end{figure}
Now in last case considering multilevel heterogeneous network we see that for DEEC, DDEEC, EDEEC and TDEEC first node dies at 1196,1262,1349,1688 rounds respectively. Tenth node dies at 1389, 1511, 1593, 2045 rounds respectively and all nodes are dead at 5547, 3999, 6734, 6734 rounds. Packets sent to the BS in DEEC, DDEEC, EDEEC and TDEEC are 106514, 79368, 236380, 314848  respectively as shown in Fig. 16, 17 and 18.
It is observed from all the above scenarios that for first case of three level heterogeneous WSN considering $a=1.5$,$b=3$,$m=0.5$ and $m_{o}=0.4$ TDEEC performs best of all, EDEEC performs better than DDEEC and DEEC where DDEEC performs better than DEEC in terms of stability period. For EDEEC and TDEEC instability period is higher as compared to DDEEC and DEEC. When values of a, b, m, $m_{o}$ are decreased linearly further in second and third scenario, same results as in first scenario are found for all protocols. In fourth and fifth scenarios when a, b, m, $m_{o}$ are increased linearly it is found after larger number of simulations that in some scenarios DEEC performs better than DDEEC, EDEEC in terms of stability period, TDEEC performs best and stability period of DDEEC and EDEEC is almost the same. Whereas instability period of TDEEC and EDEEC is also larger than DEEC and DDEEC even some nodes are not dead in EDEEC and TDEEC after 10,000 rounds. In last case considering multilevel heterogeneous network in which all nodes have random energy it is observed that TDEEC performs best of all, EDEEC performs better than DDEEC and DEEC and DDEEC performs better than DEEC in terms of stability period. For EDEEC and TDEEC instability period is higher as compared to DDEEC.
\section{Conclusion and Future Work}
We have examined DEEC, E-DEEC, T-DEEC and D-DEEC for heterogeneous WSNs containing different level of heterogeneity. Simulations prove that DEEC and DDEEC perform well in the networks containing high energy difference between normal, advanced and super nodes. Whereas, we find out that EDEEC and TDEEC perform well in all scenarios. TDEEC has best performance in terms of stability period and life time but instability period of EDEEC and TDEEC is very large. So, EDEEC and TDEEC is improved in terms of stability period while compromising on lifetime. Further research can be done on the above mentioned issue.
\bibliographystyle{plain}
\end{document}